\documentstyle[epsfig]{aipproc}
\begin{document}
{\hfill LBNL-41167}
\vskip .01 in
{\hfill December, 1997}
\vskip .01 in
\title{Bremsstrahlung and Pair Creation: \\
Suppression Mechanisms and \\ How They 
Affect EHE Air Showers}

\author{Spencer R. Klein}
\address{Lawrence Berkeley National Laboratory, Berkeley, CA, 94720} 

\maketitle
\begin{abstract}
Most calculations of air shower development have been based on the
Bethe-Heitler cross sections for bremsstrahlung and pair production.
However, for energetic enough particles, a number of different
external factors can reduce these cross sections drastically, slowing
shower development and lengthening the showers.  Four mechanisms that
can suppress bremsstrahlung and pair production cross sections are
discussed, and their effect on extremely high energy air showers
considered.  Besides lengthening the showers, these mechanisms greatly
increase the importance of fluctuations in shower development, and can
increase the angular spreading of showers.
\end{abstract}

\centerline{Presented at the Workshop on}
\centerline{\it ``Observing Giant Cosmic Ray Airshowers
for $>10^{20}$ eV Particles from Space''\rm}
\centerline{November 13-15, 1997, College Park, MD}
\vskip .2 in
\section*{Introduction}

The electromagnetic portion of high energy air showers is governed by
bremsstrahlung and pair production.  Although the formulae for these
process have been around for over 60 years, it is not well known that,
in many situations, these formulae can be very wrong.  The medium in
which the bremsstrahlung or pair production occurs can drastically
affect the cross sections.  This contribution will discuss four
different ways in which the medium can reduce the bremsstrahlung and
pair creation cross sections.

These suppression mechanisms can affect air showers by increasing the
effective radiation length, lengthening the showers, and moving the
position of the shower maximum deeper into the atmosphere.  For ground
based arrays like the proposed Auger Observatory, even a small change
in the depth of shower max can affect the energy hitting the ground,
especially for non-vertical showers where shower maximum is
considerably above ground level.  Air fluorescence detectors like Flys
Eye and the proposed OWL can measure the shower profile, and so their
energy measurement would be less affected by unforeseen changes in the
shower development.  However, a change in the position of shower
maximum can affect measurements of the composition of the highest
energy cosmic rays.  Moreover, these mechanisms will drastically
reduce the number of particles in the early stages of the shower,
changing the shower development profile.

\section{Bremsstrahlung and Pair Production Suppression Mechanisms}

Suppression mechanisms for bremsstrahlung and pair production are
possible because of the unusual kinematics in these processes.  For
ultrarelativistic particles, the momentum transfer between the
radiating electron or converting photon and the target nucleus is very
small, especially in the longitudinal direction\cite{LP}.  For
bremsstrahlung where $E\gg m$,
\begin{eqnarray}
q_\parallel & = & p_e - p_e' - k/c \\ & = & \sqrt{(E/c)^2\ -\ (mc)^2}
-\sqrt{((E-k)/c)^2\ -\ (mc)^2} - k/c = {m^2 c^3 k \over 2E (E - k) }
\label{eqlong}
\end{eqnarray}
where $p_e$ and $p_e'$ are the electron momenta before and after the
interaction respectively, $k$ is the photon energy, $m$ is the
electron mass and $\gamma=E/m$.  For ultrarelativistic electrons,
$q_\parallel$ can be very small.  For example, for a 1 EeV electron
emitting a 100 PeV photon, $q_\parallel= 10^{-9}$eV/c.  Because
$q_\parallel$ is so small, by the uncertainty principle, it must take
place over a long distance, known as the formation length:
\begin{equation}
l_{f0} = {\hbar \over q_\parallel}= {2 \hbar E (E-k) \over m^2c^3 k }.
\label{lfzero}
\end{equation}
For the above example, $l_{f0}$ is 200 meters; for a 1 PeV photon from
the same electron, $l_{f0}$ rises to 20 km.  This distance is the
distance required for the electron and photon to separate to become
distinct particles.  It is also the path length over which the
emission amplitude adds coherently to produce the emission
probability.  If something happens to the electron or nascent photon
while it is traversing the formation zone, then the coherence can be
disrupted, reducing the effective formation length and hence the
emission probability.  Even weak forces, acting over a long formation
length, can be strong enough to destroy the coherence required for
emission.  The mechanisms discussed here work by disrupting the
electron or photon, reducing the effective formation length.

\subsection{Multiple Scattering (The LPM Effect)}

Multiple scattering can cause disruption by changing the electron
trajectory.  If, taken over $l_{f0}$, the electron multiple scatters
by an angle larger than the typical bremsstrahlung emission angle
$1/\gamma$, then emission can be suppressed\cite{LP}.

The reduction can be calculated by considering the effect
multiple scattering has on $q_\parallel$; as the electron changes
direction, it's forward velocity is reduced, and, with it, producing
a change in $q_\parallel$.  This can be modelled by dividing the
multiple scattering evenly between $p_e$ and $p_e'$.  Then, 
\begin{equation}
q_\parallel = \sqrt{(E\cos{\theta_{MS/2}}/c)^2-(mc)^2}
-\sqrt{((E-k)\cos{\theta_{MS/2}}/c)^2-(mc)^2} -k/c
\label{eqlonglpm}
\end{equation}
where $\theta_{MS/2}$ is the multiple scattering in half the formation
length, $E_s/E \sqrt{l_f/2X_0}$, where $E_s=m\sqrt{4\pi\/\alpha}=21$
MeV, and $X_0$ is the radiation length.  Scattering after the
interaction is for electron energy $E-k$. This leads to a quadratic in
$l_f$:
\begin{equation}
l_f = {2\hbar E(E-k) \over km^2c^3 (1 + E_s^2 l_f /m^2 c^4 X_0)}
=l_{f0} \bigg[1  + {E_s^2 l_f \over m^2 c^4 X_0}\bigg]^{-1}.
\label{lflpm}
\end{equation}
If multiple scattering is small, this reduces to Eq.\ (\ref{lfzero}).
When multiple scattering dominates
\begin{equation}
l_f = \sqrt{2\hbar cE(E-k)X_0 \over E_s^2 k}
=l_{f0}\sqrt{kE_{LPM}\over E(E-k)}.
\label{lflpm2}
\end{equation}where $E_{LPM}$ is a material dependent constant, given by
$E_{LPM} = m^4c^7 X_0/2\hbar E_s^2 \approx 3.85\ {\rm TeV/cm}~X_0$.
For lead, $E_{LPM}=2.2$ TeV, while for water $E_{LPM}=139$ TeV and for
sea level air $E_{LPM}=117$ PeV.

Since the formation length is the maximum distance over which the
bremsstrahlung amplitude add coherently, the bremsstrahlung amplitude
is proportional to the formation length, so the suppression factor is
\begin{equation}
S= {d\sigma/dk \over d\sigma_{BH}/dk} = {l_f
\over l_{f0}} =\sqrt{kE_{LPM}\over E(E-k)}
\label{eqslpm}
\end{equation}
and the $dN/dk\sim 1/k$ found by Bethe and Heitler changes to
$dN/dk\sim 1/\sqrt{k}$.

\begin{figure}
\epsfig{file=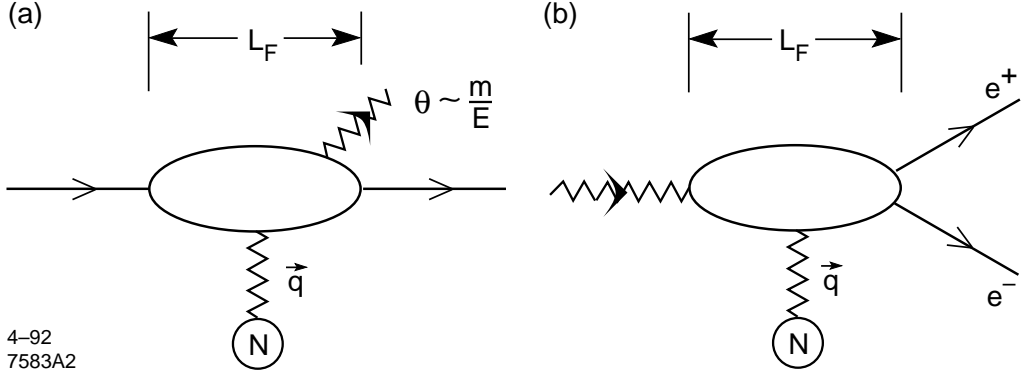,height=2in}
\caption{Schematic Representation of bremsstrahlung and pair
conversion, showing the formation zone.}
\label{schematic}
\end{figure}

A similar effect occurs for pair production, where the produced
electron and positron can multiple scatter.  The two effects are
closely related, as is shown in Fig. \ref{schematic}, and this
relationship can be used to relate the bremsstrahlung and pair
creation formation lengths and cross sections.  For
pair production
\begin{equation}
l_{f0} = {2 \hbar E  (k-E) \over m^2 c^3 k};
\label{elfpair}
\end{equation}
the corresponding suppression is 
\begin{equation}
S=\sqrt{kE_{LPM} \over E(k-E)}.
\label{espair}
\end{equation}
This calculation has several limitations.  The semi-classical approach
may fail for $k\sim E$.  The calculation neglects the statistical
nature of multiple scattering, instead treating it deterministically.
And, it neglects many niceties like the large non-Gaussian tails on
Coulomb scattering and electron-electron inelastic scattering.

Migdal developed a more sophisticated approach which avoided many of
these problems\cite{Migdal}.  He treated the multiple scattering as
diffusion, calculating the average radiation for each trajectory, and
allowing for interference between nearby collisions.  He found the
cross section for bremsstrahlung is
\begin{equation}
  {d\sigma_{\text{LPM}} \over dk} = {4\alpha r_e^2\xi(s) \over 3k}
\{y^2 G(s) + 2 [1 +(1-y)^2 ] \phi (s) \} Z^2  \ln\bigg({184 \over Z^{1/3}}
\bigg).
\label{esigmamig}
\end{equation}
where $G(s)$ and $\phi(s)$ are the solutions to differential
equations, given by\cite{Stanev}
\begin{eqnarray}
\phi(s) & = & 1-\exp{\bigg[-6s[1+(3-\pi)s]+s^3/
(0.623+ 0.796s+0.658s^2)\bigg]}   \\
\psi(s) & = & 1-\exp{\bigg[-4s-8s^2/
(1+3.96s+4.97s^2-0.05s^3+7.5s^4)} \bigg] \\
G(s) & = & 3\psi(s)-2\phi(s).
\label{estanev}
\end{eqnarray} 
where
\begin{equation}
s= {1\over 2} \sqrt{E_{LPM}k \over E(E-k)\xi(s)}.
\label{esmig}
\end{equation} 
For $k\ll E$, $s\sim 1/<\gamma\theta_{MS}>$. For $s\gg1$, there is no
suppression, while for $s\ll1$, the suppression is large.  $\xi(s)$
accounts for the increase in radiation length as photon emission
drops. Migdals solution for $s$ and $\xi(s)$ is recursive, because $s$
depends on $\xi(s)$.  The recursion can be avoided by
defining\cite{Stanev}
\begin{equation}
s'={1\over 2} \sqrt{E_{LPM}k \over E(E-k)}.
\label{esprime}
\end{equation} 
This is possible because $\xi$ depends only logarithmically on $s$; a
modified formulae for $\xi$ depends only on $s'$:
\begin{eqnarray}
\xi(s') & = & 2  \hskip 2.75 in  (s'  <   \sqrt{2}s_1) \\
\xi(s') & = & 1 + h - {0.08(1-h)[1-(1-h)^2] \over \ln{\sqrt{2}s_1}}
\hskip 0.5  in (\sqrt{2}s_1  <  s'\ll 1) \\
\xi(s') & = & 1 \hskip 2.75 in (s'  \ge  1)
\label{exiprime}
\end{eqnarray}
where $h=\ln{s'}/ln{(\sqrt{2}s_1)}$.  In the strong suppression limit,
$s\rightarrow 0$, $G(s)=12\pi s^2$ and $\phi(s)=6s$.  With these
approximations, the semi-classical $d\sigma/dk\sim 1/\sqrt{k}$ scaling
is recovered, albeit with a different coefficient.

\begin{figure}
\epsfig{file=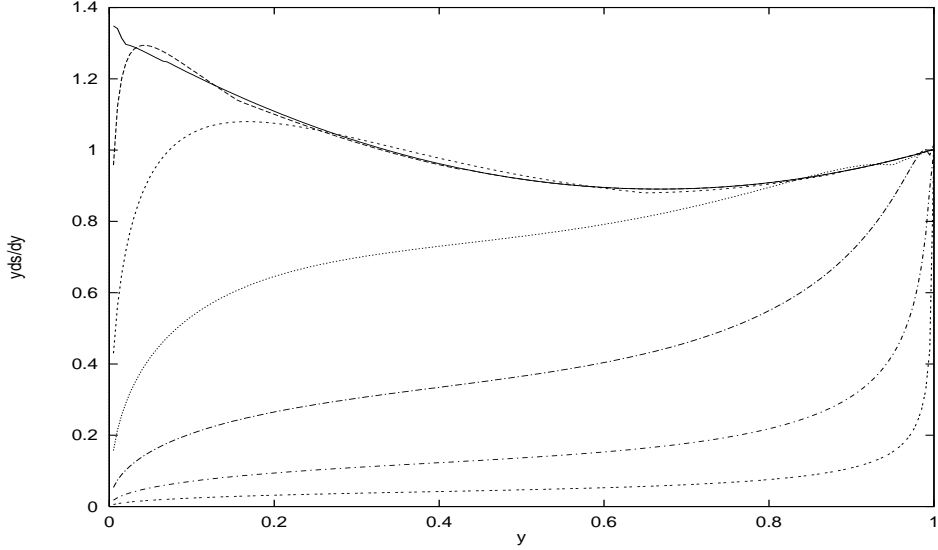,height=5in,width=3in,%
bbllx=40,bblly=50,bburx=570,bbury=760,%
clip=,angle=270}
\caption{$yd\sigma_{\text{LPM}}/dy$ for bremsstrahlung as a function
of $y$ for various electron energies in lead ($E_{LPM}$=2.2 TeV),
showing how the spectral shape changes.  Electrons of energies 10 GeV,
100 GeV, 1 TeV, 10 TeV, 100 TeV, 1 PeV and 10 PeV are shown.  The
upper curve (solid line) is for $E=10$ GeV, and the emission drops as
energy rises.}
\label{brem}
\end{figure}

Fig. \ref{brem} shows how the LPM effect reduces $yd\sigma/dy$
($y=k/E$) for electrons in a lead target.  The 10 GeV electron curve
is very close to the Bethe-Heitler prediction; in the absence of
suppression, this curve would hold for all electron energies.  As the
electron energy rises, emission drops.  At the highest electron
energies, photons with $k\ll E$ are almost completely suppressed.  For
$E^2/E_{LPM} < k < 1.3 E^2/E_{LPM}$, the Migdal curve rises slightly
above the unsuppressed; this is a consequence of either the approach
or the approximations Migdal used.

\begin{figure}
\epsfig{file=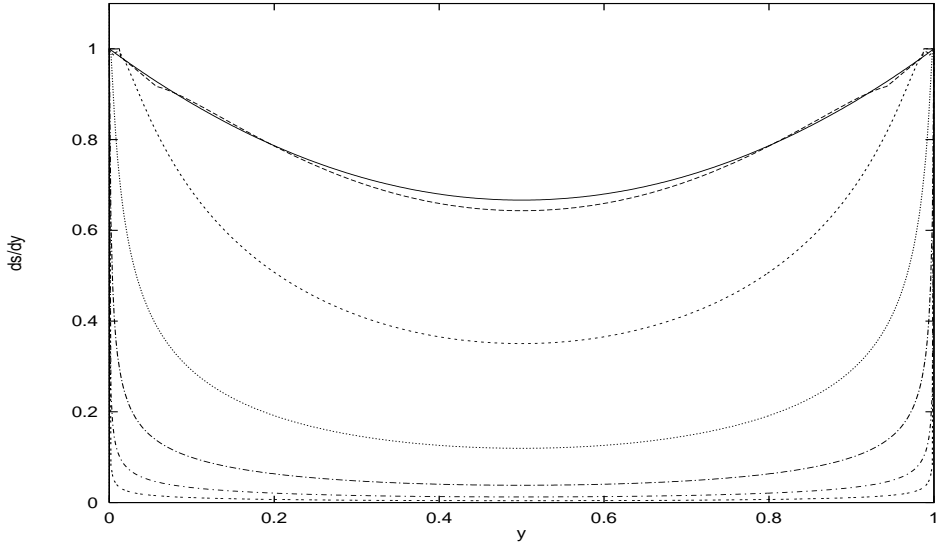,height=5in,width=3in,%
bbllx=40,bblly=50,bburx=570,bbury=760,%
clip=,angle=270}
\caption{$d\sigma_{\text{LPM}}/dy$ for pair production in lead, as a
function of $y$ for various photon energies, showing how the spectral
shape changes.  The solid line is for $k=1$ TeV, and the cross
sections drop with energy; the other curves are for photons of
energies 1 TeV, 10 TeV, 100 TeV, 1 PeV, 10 PeV, 100 PeV and 1 EeV.}
\label{pair}
\end{figure}

Fig. \ref{pair} shows how the pair production cross section is reduced.
Compared with bremsstrahlung, pair production suppression sets in at
higher energies.  Symmetric pairs are suppressed the most; in the
extremely high energy limit, one of the produced electrons takes
almost all of the photon energy.  So, where the LPM effect is
extremely strong, an electromagnetic shower becomes a succession of
interactions where an electron emits a bremsstrahlung photon that
takes almost all of the electrons energy, followed by a very
asymmetric pair conversion, producing an electron or positron with
almost all of the energy of the initial lepton.

\begin{figure}
\epsfig{file=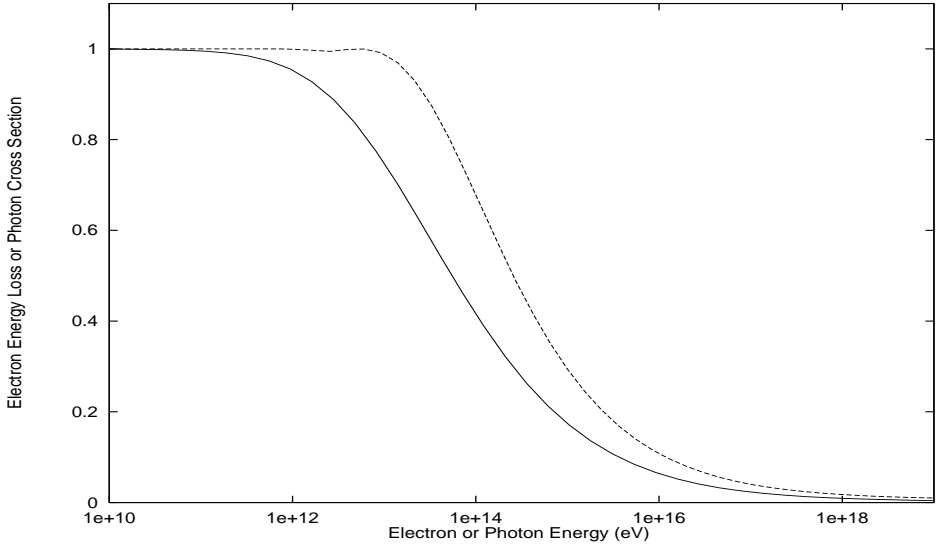,height=5in,width=3in,%
bbllx=40,bblly=50,bburx=570,bbury=760,%
clip=,angle=270}
\caption{The reduction in electron energy loss ($\int_0^{E} (kdN/dk)
dk$) due to the LPM effect (solid line) and in the photon conversion
cross section (dashed line) due to the LPM effect in a lead target.
The LPM effect turns on at higher energies for photons.}
\label{compare}
\end{figure}

Two metrics for suppression effects in showers are the electron energy
loss ($dE/dx$) and photon conversion cross section.
Fig. \ref{compare} shows how these two markers change with energy in a
lead target.  As the particle energy rises above $E_{LPM}$ (2.2 TeV
for lead), the energy loss and conversion cross section fall, with
bremsstrahlung affected at lower energies than pair conversion.
 \def\xo{$X_0$}

\begin{figure}
\epsfig{file=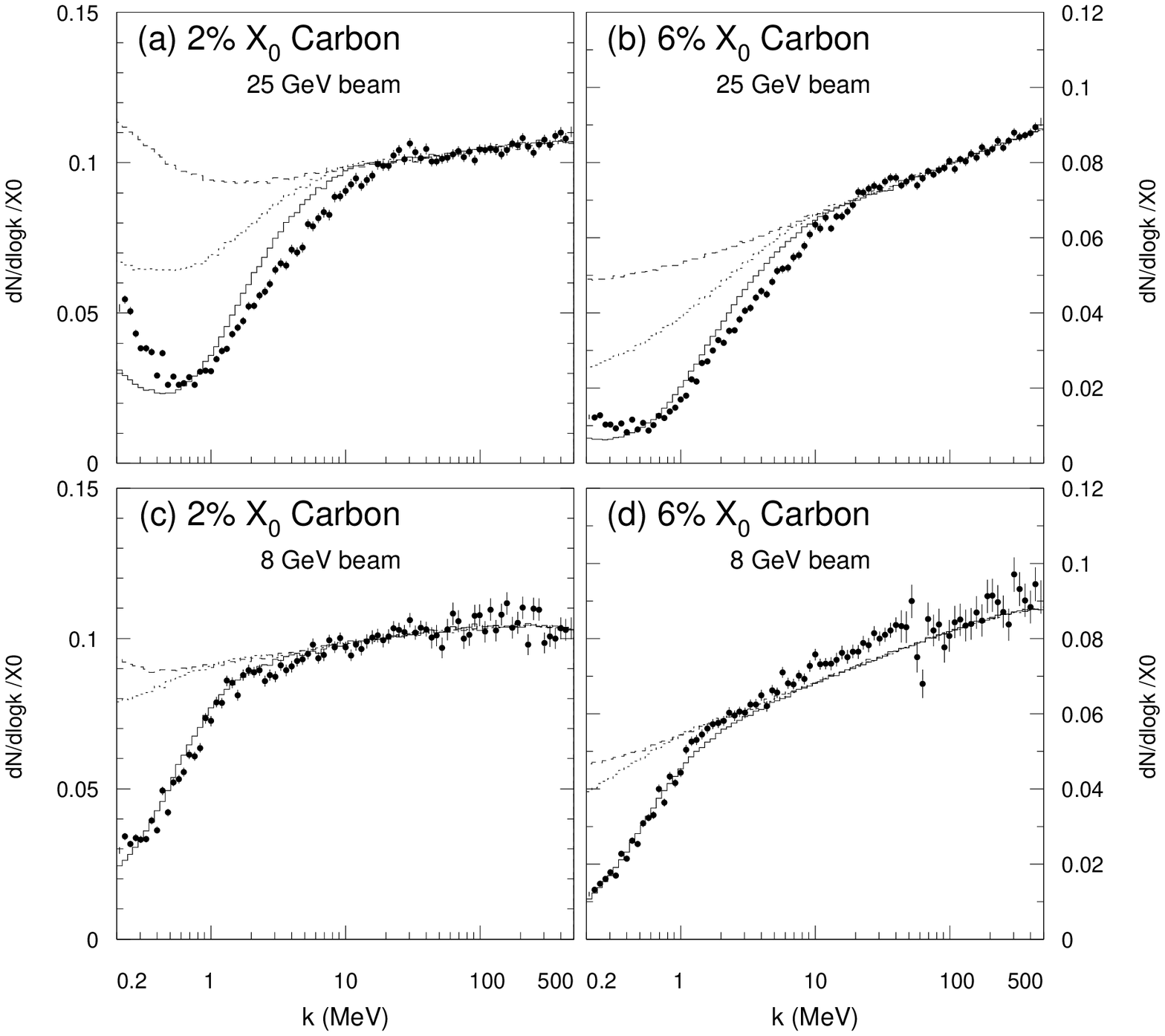,height=4.5in,width=5.5in,%
bbllx=29,bblly=160,bburx=566,bbury=650,%
clip=,}
\caption{Comparison of data from SLAC-E-146 with Monte Carlo
predictions for 200 keV to 500 MeV photons from 8 and 25 GeV electrons
passing through 2\% and 6\% $X_0$ thick carbon targets. The cross
sections are $dN/d(\log k)/X_0$ where $N$ is the number of photons per
energy bin per incident electron, for (a) 2\% \xo carbon and (b) 6\%
\xo carbon targets in 25 GeV electron beams, while (c) shows the 2\%
\xo carbon and (d) the 6\% \xo carbon target in the 8 GeV beam.  Monte
Carlo predictions are shown for LPM plus dielectric suppression (solid
line), LPM suppression only (dotted line) and Bethe-Heitler (dashed
line); all include transition radiation.}
\label{carbon}
\end{figure}

The LPM effect was studied experimentally by the SLAC E-146
collaboration, who sent 8 and 25 GeV electrons through thin (0.07 \%
to 6\% of $X_0$) targets of materials ranging from carbon to gold.
Fig. \ref{carbon} shows their data for carbon\cite{E146}, the material
that is closest to air.  Both LPM and dielectric suppression
are needed to explain the data.

Since E-146, there have been a number of additional calculations of
bremsstrahlung with multiple scattering, using a variety of different
approaches\cite{Bland,Zak,French,Baier}.  Several calculations have
showed that, the non-Gaussian tail of large angle Coulomb scatters
introduces additional term into the cross section; in the limit of
large suppression\cite{Zak,French,Baier}
\begin{equation}
S= \sqrt{{kE_{LPM}\over E(E-k)}\log{\big({E(E-k) \over kE_{LPM}}\big)}}
\label{eqslpm2}
\end{equation}
This will reduce the magnitude of suppression in extreme conditions.
Inelastic interactions, involving the atomic electrons should be
treated separatedly from the elastic nuclear interactions, and a
different potential should be used\cite{Zak,Baier}.  This may explain
the poor agreement observed by E-146 between the data and the Migdal
calculations for their light carbon and aluminum targets, especially
for the 25 GeV data, around $k=E^2/E_{LPM}$.  In light of these
results, calculations based on Migdal's formulae should be treated
with caution, especially for light targets like air, where the
inelastic form factor is important.

\subsection{Dielectric Suppression}

Dielectric suppression occurs because photons produced in
bremsstrahlung can interact with the atomic electrons in the medium by
forward Compton scattering\cite{Term}.  The photon wave function
acquires a phase shift depending on the dielectric constant of the
medium $\epsilon(k)=1-(\hbar\omega_p)^2/k^2$ where $\omega_p$ is the
plasma frequency of the medium.  With this substitution, the photon
momentum ($k/c$) in Eq. (1) becomes $\sqrt\epsilon k/c$, and
$q_\parallel$ acquires an additional term $(\hbar\omega_p)^2/2ck$.
This leads to a reduced $l_f$, and a suppression factor
\begin{equation}
S = { k^2 \over k^2 + (\gamma\hbar\omega_p)^2 }.
\end{equation}
This effect only applies for small $y$, $y<y_{die}=\omega_p/m$.  For
typical solids, $\omega_p\sim 60$ eV, so $y_{die}\sim 10^{-4}$.  For
$y<y_{die}$, the photon spectrum becomes $d\sigma/dk \sim k$,
suppressing the emission by $(\gamma\hbar\omega_p)^2/k^2$.  The effect
is also clealy apparent in the E-146 data in Fig. \ref{carbon}.

\subsection{Suppression of Bremsstrahlung by Pair Creation, and vice-versa}

As Landau and Pomeranchuk pointed out, when $l_{f0}$ becomes as long
as $X_0$, then partially created photons can be pair create,
destroying the coherence\cite{LP}.  A simple calculation of this
effect can be done by limiting $l_f$ to a maximum of $1
X_0$\cite{Galitsky}.  This suppression is visible (stronger than LPM
and dielectric suppression) for
\begin{equation}
E>E_p = {2 X_0 \omega_p E_s\over \hbar c}.
\end{equation}
For $E>E_p$, there is a 'window' in $k$ where this mechanism applies:
$k_{p-}=X_0\omega_p^2/2\hbar c < k < k_{p+}=2\hbar
cE(E-k)/(X_0E_s^2)$. In this region, the photon spectrum is suppressed
by $k$, and $d\sigma/dk$ is constant. For $k<k_{p-}$, dielectric
suppression dominates, while for $k>k_{p+}$, the LPM effect is
dominant.  $E_p$ is 1.6 PeV in water or ice, and 42~PeV for air at sea
level.  The $E_p$s are so similar because the variation in $X_0$ and
$\omega_p$ partially offset each other.  In sea level air, the
'window' is 331 MeV$<k< 3.0\times10^{-24} E^2$ (with $E$ in eV).

There should be a similar effect where a partially produced electron
or positron emits a bremsstrahlung photon.  This possibility limits
the coherence over $l_{f0}$.  The strength of this suppression has not
yet been calculated, but it should be comparable to that for pair
production suppressing bremsstrahlung.

This approach is overly simplistic. The 'hard cutoff' in $l_f$ should
be replaced by a probabilistic approach, where the photon interaction
probability depends on the distance travelled.  And, other suppression
mechanisms will also be in effect, the radiation length will be longer
than the naive $1 X_0$.  An accurate calculation should include the
interplay between the two reactions to arrive at an overall effective
shower distance.  Still, the above equation gives a reasonable
estimate of when this effect needs to be considered.

\begin{figure}
\epsfig{file=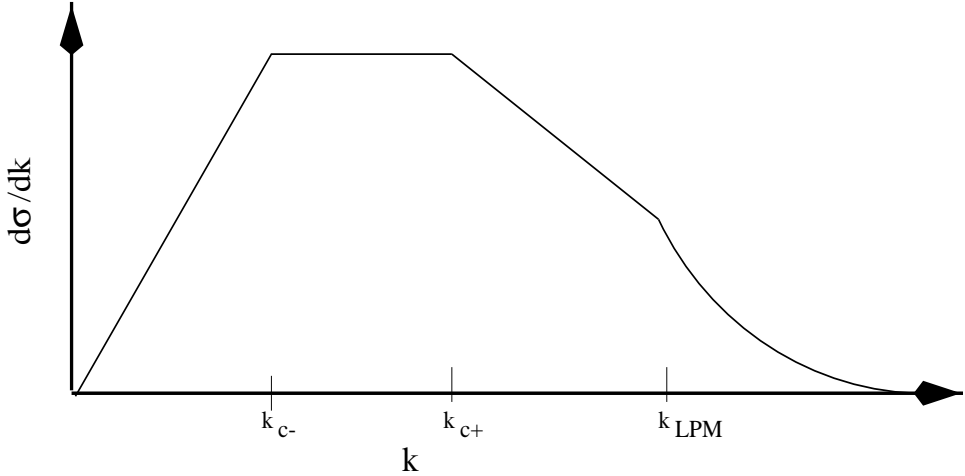,height=2.5in,width=5in,%
bbllx=0,bblly=0,bburx=520,bbury=247,%
clip=,angle=0}
\caption{Schematic view of bremsstrahlung $d\sigma/dk$ when several
suppression mechanisms are present.  Below $k_{c-}$, dielectric
suppression dominates.  Between $k_{c-}$ and $k_{c+}$, suppression due
to pair creation is dominant.  Between $k_{c+}$ and
$k_{LPM}=E^2/E_{LPM}$, the LPM effect is most important, while above
$k_{LPM}$ the Bethe-Heitler spectrum is present.  For $E<E_p$, pair
creation suppression disappears and LPM suppression connects with
dielectric suppression.}
\label{ssummary}
\end{figure}

Fig. \ref{ssummary} summarizes these results, and shows that a
'simple' bremsstrahlung photon spectrum can have several different
slopes for different photon energies.

\subsection{Magnetic Suppression}

An external magnetic field can also suppress bremsstrahlung and pair
creation.  The bending caused by an external field acts just as does
multiple scattering.  The difference is that the magnetic field
bending is quite deterministic, while the scattering angles are
randomly distributed.  The magnetic bending angle is $\theta_B =
\Delta p/p = eB l_f \sin{(\theta_B)}/E$ where $B$ is the magnetic
field and $\theta_B$ is the angle between the field and the electron
trajectory.  Emission is reduced if $\theta_B > 1/\gamma$; this
happens for $y<2\gamma B\sin{(\theta_B)}/B_c$, where
$B_c=m^2c^3/\hbar$ is the critical field strength\cite{klein}.

The magnitude of the suppression can be found using an approach
similar to that used by Landau and Pomeranchuk for multiple
scattering.  Because $\theta_B\sim l_f$, while
$\theta_{MS}\sim\sqrt{l_f}$ the energy dependence is different.  With
the definition $E_B=mB_c/B\sin{(\theta_B)}$, the suppression factor
is\cite{klein2}
\begin{equation}
S=\bigg[ { kE_b \over E(E-k)}\bigg]^{2/3}.
\end{equation}
A magnetic field will also bend pair produced electrons and positrons.
The same equation holds, except that $k-E$ replaces $E-k$.  For a
particle trajectory perpendicular to Earth's $\sim 0.5$ Gauss field,
$E_B\sim 45$ EeV.  So, for cosmic rays with energies above $10^{20}$
eV, magnetic suppression must be considered.

\subsection{Emission Angles}
\label{angle} 

Besides reducing the interactions rates, these mechanisms also
increase the angular spread of electromagnetic showers.  When
bremsstrahlung and pair creation occur with a large opening angle,
the formation length is shortened.  For bremsstrahlung
\begin{equation}
l_{f0}={ 2\hbar E(E-k) \over m^2c^3k (1+\gamma^2\theta_\gamma^2)}
\end{equation}
where $\theta_\gamma$ is the angle between the photon and the electron
trajectory.  When $\theta_\gamma$ is included in the calculations, the
average emission angle rises from $\theta_\gamma\sim 1/\gamma$ to
$\theta_\gamma\sim 1/S\gamma$.  If $S$ is small enough, this spreading
can dominate over multiple scattering in determining the angular
spreading of the shower.  If the scattering is taken over $1/2X_0$
(assuming 1 interaction per $X_0$, and half the particles are charged
and hence subject to multiple scattering), this occurs for roughly
$S<0.05$.  Most calculations of air and water showers do not include
this angular spreading.  It is likely to be most important for
calculations for radio emission from showers in air and ice\cite{Zas}.

\section{Suppression in Showers}

Suppression mechanisms can affect showers by slowing their
development, effectively increasing the radiation length.  Because
they all work by reducing $l_f$, the effects do not add; instead,
$q_\parallel$ must be summed, and the total suppression
calculated. Usually, multiple scattering is the most significant cause
of suppression, so it will be the focus of this section.

The effective increase in radiation length due to multiple scattering
can be seen in Figs. \ref{compare}; it shows how the area under the
curve in Figs. \ref{brem} and \ref{pair} drops as the incident
particle energy rises.  For bremsstrahlung, energy loss is halved for
electrons with $E=22E_{LPM}$, while the pair production cross section
is halved for $k=100E_{LPM}$.  Then, the radiation length in
the first generation of the shower doubles; succeeding generations
will show smaller effects.

However, the effects go beyond simply lengthening showers.  Because
soft photon bremsstrahlung is the first reaction to be affected, the
number of interactions will decrease more rapidly then the electron
energy loss.  So, the initial part of a shower will consist almost
entirely of a few high energy particles, without an accompanying
'fuzz' of lower energy particles.  The initial shower development
depends on a much smaller number of interactions.  For example, a 25
GeV electron in sea level air will emit about 14 bremsstrahlung
photons per $X_0$, while a $10^{17}$ eV electron will emit only 3.  It
is worth noting that, these numbers are naturally finite because
dielectric suppression eliminates the infrared divergence.  Pair
production is similar; the pairs become increasingly asymmetric. The
higher energy lepton from a 25 GeV photon pair conversion takes an
average of 75\% of $k$; for a $10^{19}$ eV photon, the average is more
than 90\%.

Because of this, shower to shower fluctuations become much larger.
Misaki studied shower development in lead and standard rock.  For
$E\gg E_{LPM}$ he found that the position of shower maximum was
shifted to larger depths, and that the position of shower maximum
varied greatly from shower to shower, and that the shower to shower
variations overshadowed the average shower development\cite{Misaki}.

In the limit $E\gg E_{LPM}$, the initial part of an air shower becomes
a succession of asymmetric pair production, where the higher energy of
the pair loses most of it's energy to a single bremsstrahlung photon,
re-starting the process.  In short, the paradigm that successive
generations of air showers have twice as many particles with half as
much energy as the current generation fails completely.

\section{Air Showers}

The composition of the highest energy cosmic rays is still a mystery.
This article will consider two possibilities: protons (the most
popular) and photons.  In both cases, we will take $3\times10^{20}$ eV
as a standard energy. For incoming heavy ions, the effects are greatly
reduced because of the lower per particle energy, and neutrons can be
treated as protons.  Because these are toy models, the likely
possibility of the photon pair converting in the earth's magnetic
field will be neglected.

Most current works have considered proton initiated showers.  There
has been disagreement as to whether the LPM effect is important in air
showers.  Capdevielle and Atallah found that it had a large effect on
$10^{19}$ eV and $10^{20}$ eV showers\cite{Cape}.  However, Kalmykov,
Ostapchenko and Pavlov found a much smaller effect; for a $10^{20}$ eV
incident proton, the number of electrons at shower maximum decreased
by 5\%, while the position of shower maximum shifted downward by
$15\pm2$ g/cm$^2$\cite{KOP}.  This shift is less than the error on a
typical measurement of shower maximum in a single shower, but it can
have a significant effect on composition studies.  Of course, as
experiments probe higher energies, suppression becomes more important.
None of these authors considered the effects of fluctuations or other
suppression mechanisms.

Air showers studies are complicated by the fact that density, and
hence $E_{LPM}$ and $E_p$ depend on altitude and temperature.
Ignoring temperature changes, pressure decreases exponentially with
altitude, with scale height 8.7 km.  For showers, it is convenient to
work with column depth measured in g/cm$^2$.  In an isothermal model,
then $E_{LPM}= 117$ PeV$(A_0/A$), where $A$ is the column depth and
$A_0$ is ground level, 1030 g/cm$^2$.  Temperature will modify the
relationship; with a temperature correction this $E_{LPM}$ is 2.25 EeV
at 36 g/cm$^2$ (1 $X_0$) depth, and 1 EeV at 90 g/cm$^2$ (1 hadronic
interaction length, $\Lambda$).  Neglecting temperature,
$y_{die}=1.3\times10^{-6}A/A_0$.  For pair creation suppression,
$E_p=42$ PeV$\sqrt{(A_0/A)}$.  The corresponding photon window is 331
MeV $< k < 3.0\times10^{-24}E^2 (A_0/A)$.  The two window
'edges' have different $A$ dependencies because all three mechanisms
have a different dependence on $\omega_p$ and $X_0$.

Incoming photons react by pair production, while protons interact
hadronically.  A central hadronic collision will produce a shower of
several hundred pions; the neutral pions will decay to photons.  The
highest energy $\pi^0$ will have a rapidity near to the incoming
proton, and their decay photons will have energies around
$2\times10^{19}$eV.  Many diffractive processes, such as $\Delta$
production can produce photons with similar energies.  Overall,
photons from central interaction will have an average energy of about
$2\times10^{17}$ eV.

\begin{figure}
\epsfig{file=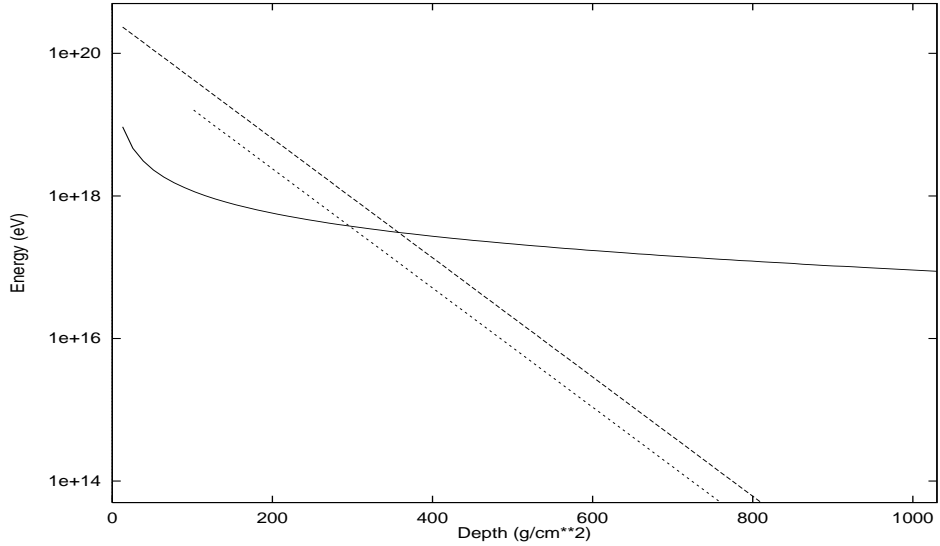,height=5in,width=3in,%
bbllx=40,bblly=50,bburx=570,bbury=760,%
clip=,angle=270}
\caption{$E_{LPM}$ (solid line), average particle energy for a
$3\times10^{20}$ eV photon shower (dashed line) and average particle
energy for a $2\times10^{19}$eV photon created at $1\Lambda$ (short
dashes).  At altitudes where the average particle energy is larger
than $E_{LPM}$, suppression is important; with the ratio of the two
energies determining the degree of suppression.}
\label{airdepth}
\end{figure}

Although a complete Monte Carlo simulation is required to understand
the effects of suppression in air showers, simple calculations can
provide some indications where it matters, and should differentiate
between the results of Capdevielle and Atallah and Kalmykov and
collaborators.  Because $E_{LPM}$ decreases with depth, in concert
with the average particle energy, suppression mechanisms can actually
become stronger as one moves deeper in the atmosphere.  The solid
curve in Fig. \ref{airdepth} shows $E_{LPM}$ as a function of
altitude.  The dashed curve shows the average particle energy, for an
idealized Bethe-Heitler shower from a $3\times10^{20}$ eV photon.  In
each successive radiation length, there are twice as many particles
with half the energy.  The curve with the short dashes shows a similar
cascade, from a $2\times10^{19}$ eV photon starting at $1\Lambda$.
The electromagnetic interactions at a given depth are determined by
the ratio of the two curves, which gives $E/E_{LPM}$.  For the photon
case, the maximum suppression occurs around 75 g/cm$^2$, where $E\sim
80 E_{LPM}$.  Electron $dE/dx$ is reduced about 80\%, and pair
production cross section is down by 60\%; the radiation length has
more than doubled. For the hadronic case, the effect is smaller, and,
of course, these high energy photons are only a small portion of the
total shower.  On the other hand, hadronic interactions are only
partially inelastic, and the proton may carry a significant fraction
of it's momentum deeper in the atmosphere, where suppression is
larger.  Because of the large variations in energy deposition depths,
it is difficult to give more quantitative estimates.

This model underestimates the effect of suppression, because, with
suppression shower development is slower than the idealized model,
further increasing the amount of suppression in the next stage.
However, it does show that suppression is important in photon shower,
and in at least parts of proton initiate showers.  The effect is
clearly smaller than that predicted by Capdevielle and Atallah, but is
consistent with Kalmykov and collaborators.

Beyond the affect on average showers, fluctuations must be
considered. because the cosmic ray energy spectrum falls as
$dN/dE\approx 1/E^3$, it is important to understand the tails of the
energy resolution distribution; without accurate simulations, showers
whose energy is overestimated can easily skew the measured spectrum.

Fluctuations can affect both ground based arrays as well as air
fluorescence detectors that optically measure the shower development.
For ground based arrays, suppression can change the relative position
of shower maximum and the detector array. Although a sea level
detector is near shower maximum for $10^{20}$ eV vertically incident
proton showers, for non-vertical showers, where the detector is
deeper, it may be significantly behind shower maximum.  The angular
spreading discussed in Subsection \ref{angle} must also be considered.

Air fluorescence detectors will observe far fewer particles in the
early stages of the shower than current simulations indicate.  LPM,
dielectric and pair creation suppression must all be considered.
Although LPM suppression affects the widest range of energies, the
other mechanisms produce a larger reduction in cross section where
they operate.  For $E> E_p$, for example, dielectric suppression will
reduce the number of bremsstrahlung photons with $k<331$ MeV by two
orders of magnitude. Pair creation suppression is more energy
dependent, but, for example, at 200 g/cm$^2$ depth, emission of
$10^{10}$ eV photons from $10^{16}$ eV electrons will be reduced by a
factor of 10.  Monte Carlo simulations are required to find the actual
shower profile, but the initial stages of the shower will be much less
visible than current simulations indicate; neither the profiles
produced by Capdevielle and Atallah nor by Kalmykov will be accurate.

Even relatively late in the shower, there will be a reduction in the
number of low energy particles.  For example, where the average
particle energy is $\sim10^{13}$ eV, LPM suppression will reduce the
number of photons below $\sim$500 MeV.  So, there can be measurable
effects even relatively close to shower maximum.  if a particle
counting detector is too far above shower maximum, then it may
underestimate the size of the shower.

\begin{figure}
\epsfig{file=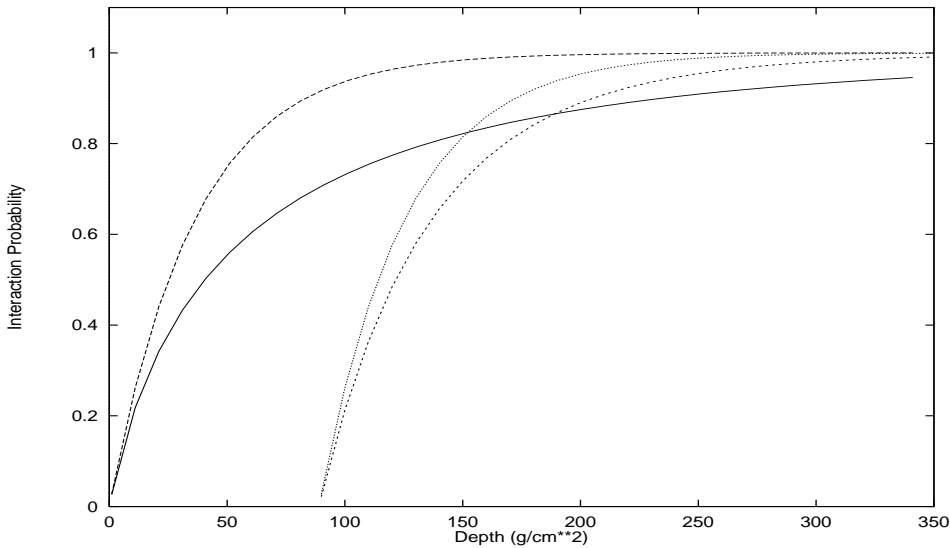,height=5in,width=3in,%
bbllx=40,bblly=50,bburx=570,bbury=760,%
clip=,angle=270}
\caption{Interaction probability for photons from cosmic rays, as a
function of depth in the atmosphere.  The solid and dashed line show
the interaction probability for a $3\times10^{20}$eV photon incident
on the top of the atmosphere for LPM and Bethe-Heitler interaction
probabilities respectively.  The short dashed and dotted lines show
the interaction probability for a $2\times10^{19}$ eV photon
produced in an interaction at a depth of $1 \Lambda$, also
for LPM and Bethe-Heitler interactions respectively.}
\label{probconvert}
\end{figure}

Another way to visualize how suppression works is to consider the
probability of photons penetrating deep into the atmosphere.  Because,
for a given photon energy, suppression increases with depth, high
energy photons have a non-negligible probability of penetrating deep
into the atmosphere.  This can drastically change the development of
photons showers.  For proton showers, it can create small, dense
subshowers within the main shower.  Fig. \ref{probconvert} shows the
interaction probability as a function of depth for a $3\times 10^{20}$
eV vertically incident photon.  The solid and dashed curves show the
LPM and Bethe-Heitler cases respectively.  With Bethe-Heitler
interaction probabilities, essentially all of the photons have
interacted by 3$X_0$ in depth, while with LPM interaction
probabilities, the photon has a 7\% chance of surviving to 10$X_0$.

For protons, the effect is smaller, The other curves in Fig.
\ref{probconvert} are for a $2\times10^{19}$ eV photon, produced by a
hadronic interaction at $1\Lambda$ in depth.  With Bethe-Heitler
(short dashes), almost all have interacted by 6 $X_0$, while with LPM
(dots) cross sections, it has a $1.2\%$ chance of surviving to
10$X_0$.  The effect on the average shower is small, but, because of
the steeply falling spectrum, it is necessary to consider even
relatively atypical showers.

Not considered here is magnetic suppression, which can dominate over
other mechanisms for interactions in the upper 3 g/cm$^2$ of the
atmosphere.

\section{Conclusions and Acknowledgements}

Multiple scattering, the dielectric of the medium, bremsstrahlung and
pair production themselves and external magnetic fields can suppress
bremsstrahlung and pair production, with bremsstrahlung of low energy
photons the most subject to suppression.  Suppression also broadens
the angular spread of emitted bremsstrahlung photons and produced
pairs.

For energetic enough particles, these mechanisms reduce electron
$dE/dx$ and photon pair production cross sections.  When this happens,
electromagnetic showers are lengthened, and shower to shower
fluctuations become much larger.  When suppression gets very large,
then the shower angular development can be dominated by interactions,
rather than multiple scattering.

The importance of suppression in air showers depends on the incident
particle type and energy.  If the incoming particles are photons, then
at $3\times10^{20}$ eV, the effects are large.  For $3\times10^{20}$
protons, the effects on the average shower is smaller, but the effects
of fluctuations may be considerable, particularly when particular
detector are included.  As experiments probe to higher energies, of
course, the effects will grow larger.

The calculations presented here show the need for a complete Monte
Carlo, including all relevant suppression mechanisms, in order to
correctly determine the shower profile for showers with energies much
above $10^{20}$ eV.

I would like to thank my E-146 collaborators for many useful
discussions.  This work was supported by the USDOE under contract
number DE-AC-03-76SF00098.

\end{document}